\documentclass[epj-guido]{svjour-guido}
\usepackage{graphics,color}
\begin{document}
\title{Comment on "Missing Transverse-Doppler Effect in Time-Dilation Experiments with High-Speed Ions"}
\author{G. Saathoff\inst{1} \and  S. Reinhardt\inst{1} \and R.
Holzwarth\inst{1} \and T.W. H\"ansch\inst{1} \and Th. Udem\inst{1}
\and D. Bing\inst{2} \and D. Schwalm\inst{2} \and A. Wolf\inst{2}
\and S. Karpuk\inst{3} \and G. Huber\inst{3}  \and C.
Novotny\inst{3,4,5} \and B. Botermann\inst{4} \and C.
Geppert\inst{4,5} \and W. N\"ortersh\"auser\inst{4,5} \and T.
K\"uhl\inst{5} \and T. St\"ohlker\inst{5}
\and G. Gwinner\inst{6}
}                     
%
%
\institute{Max-Planck-Institut f\"ur Quantenoptik, 85748 Garching,
Germany \and  Max-Planck-Institut f\"ur Kernphysik, 69029
Heidelberg, Germany \and  Institut f\"ur Physik, Universit\"at
Mainz, 55099 Mainz, Germany \and Institut f\"ur Kernchemie,
Universit\"at Mainz, 55099 Mainz, Germany \and Helmholtzzentrum
f\"ur Schwerionenforschung GSI, 64291 Darmstadt, Germany  \and Dept.
of Physics \& Astronomy, University of Manitoba, Winnipeg R3T 2N2,
Canada}
\date{}
%
\abstract{In an article "Missing Transverse-Doppler Effect in Time-Dilation Experiments with High-Speed Ions" by
S. Devasia [arXiv:1003.2970v1], our
recent Doppler shift experiments on fast ion beams are reanalyzed.
Contrary to our analysis, Devasia concludes that our results provide
an "indication of Lorentz violation". We argue that this conclusion
is based on a fundamental misunderstanding of our experimental
scheme and reiterate that our results are in excellent agreement
with Special Relativity.
%
} 
\authorrunning{ }
\titlerunning{ }
\maketitle
%
We have performed experiments of the Ives-Stilwell (IS)
type~\cite{IS1938} that test time dilation of Special Relativity
(SR) via the relativistic Doppler
shift~\cite{Saathoff2003,Reinhardt2007,Novotny2009,Botermann2011}. A
beam of ions, which exhibit an optical transition with a frequency
$\nu_0$ in their rest frame, is stored at velocity $\beta=v/c$ in a
storage ring. To resonantly excite these ions by a laser at rest in
the laboratory frame, the frequency $\nu$ of the laser needs to be
Doppler shifted according to $\nu = \nu_0 / \gamma (1-\beta \cos
\theta)$, where $\theta$ is the angle between the laser and the ion
beam, measured in the laboratory frame, and $\gamma$ governs time
dilation. For a parallel {$(\theta_{p} = 0)$ or an antiparallel
$(\theta_{a} = \pi)$} laser beam the frequencies required are
$\nu_{p,a} = \nu_0 / \gamma (1 \mp \beta)$, respectively.
Multiplying these two frequencies and using $\gamma =
(1-\beta^2)^{-1/2}$ as predicted by SR results in
\begin{equation}
\nu_p \nu_a / \nu_0^2 =1, \label{eq:1}
\end{equation}
i.e. the geometric mean of the Doppler shifted frequencies equals
the rest frame frequency for all velocities $\beta$.

In one of our implementations of the IS experiment saturation
spectroscopy is used by overlapping simultaneously a parallel and
antiparallel laser beam with the ion beam to select a narrow
velocity class $\beta_0$ within the ions' velocity distribution. The
parallel laser is held fixed at the laser frequency $\nu_{p} = \nu_0
/ \gamma (1 - \beta_0)$ and is resonant with ions at $\beta_0$,
while the other laser is scanned over the velocity distribution. The
fluorescence yield, measured with a photomultiplier (PMT) located
around 90 degree with respect to the ion beam, will exhibit a
minimum (a Lamb dip)
\begin{figure}[top]
\resizebox{0.43\textwidth}{!}{
\includegraphics{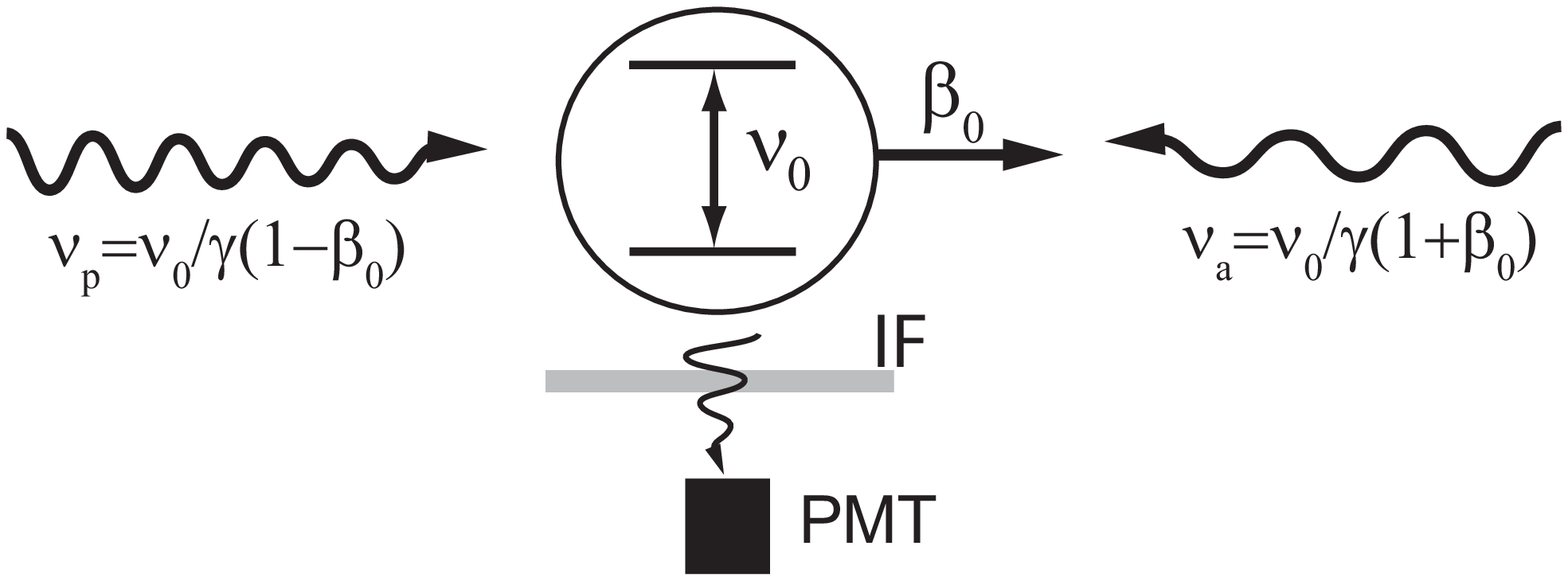}
}
\label{fig:1}       
\end{figure}
when the antiparallel laser talks to the same velocity class
$\beta_0$, i.e. when its frequency is at
$\nu_a=\nu_0/\gamma(1+\beta_0)$. SR thus predicts the Lamb dip to
occur when Eq.~\ref{eq:1} is fulfilled, which is shown to be
confirmed by our experiments to an accuracy of $<2\times 10^{-10}$
on Li{$^+$} ions at $\beta_0=0.03$ and
{$\beta_0=0.06$}~\cite{Reinhardt2007}.

S. Devasia~\cite{Devasia2010} claims that the Doppler shift of the
emitted light has to be taken into account and replaces $\nu_0$ in
Eq.~\ref{eq:1} by $\gamma\nu_0$, i.e. by the frequency of the light
detected exactly at $\theta=\pi/2$. This is a misconception of our
experimental measurement scheme. While it is true that the detected
light is Doppler-shifted, this Doppler shift is irrelevant for the
analysis. Neither do we measure the frequency of the emitted light
nor do we intend to observe at exactly right angle. We only record
the number of re-emitted photons as a function of the scanning laser
frequency to monitor the Lamb dip caused by the simultaneous
resonance of both lasers with the same ions. Thus the angle of
detection is irrelevant but $\theta\approx\pi/2$ helps to separate
fluorescence from laser stray light. In fact, stray light
suppression is the only reason for using an interference filter (IF)
in front of the PMT; its transmission width of {$10$~nm} corresponds
to {10 THz}, about $10^6$ times broader than the width of the Lamb
dip, and a factor of {10} larger than the transverse Doppler shift
(at $\beta=0.064$). None of the filters employed in our
experiments~\cite{Saathoff2003,Reinhardt2007,Novotny2009,Botermann2011}
to improve the signal-to-noise ratio in the fluorescence light
detection are affecting the shape and position of the signal
indicating the resonance of the parallel and antiparallel laser with
the same velocity class $\beta_0$.  The frequency $\nu_0$ occurring
in Eq.~\ref{eq:1} has nothing to do with the frequency of the
emitted light in our experiment, but is the rest frame frequency
$\nu_0$ deduced from experiments at smaller ion
velocities~\cite{Reinhardt2007,Riis1994}.

In conclusion, SR predicts Eq. ~\ref{eq:1} as the outcome of our
experiments, which is confirmed with high accuracy.

%

\end{document}